\documentclass[traditabstract]{aa} % for the abstract without structuration
\usepackage{graphicx}
\usepackage{txfonts}
\begin{document}
\title{A New Possible Way to Explain the DAMA Results}

\author{J.Va'vra
\inst{}
}

\institute{SLAC, Stanford University, CA94309, U.S.A.\\
\email{jjv@slac.stanford.edu}
}

\date{Received January 10, 2014}

% \abstract{}{}{}{}{} 
% 5 {} token are mandatory

\abstract
% context heading (optional)
% {} leave it empty if necessary  
{At present there is an effort to reconcile the results of the DAMA experiment with those from other Dark Matter experiments such as CoGeNT, CRESST, 
CDMS, and all LXe experiments. The author suggests a new model describing the Dark Matter signal as the result of collisions of very 
light (1-to-few~GeV/$c^2$) WIMPs with hydrogen, and compares it with currently accepted models of collisions with heavy nuclei (Na, Ge or Xe). 
The hydrogen target would come from H-contamination of NaI(Tl), Ge and CaWO$_{4}$ crystals. Initial tuning indicates that one can 
explain the modulation amplitude of DAMA and CoGeNT with this model, assuming a WIMP-proton cross section between $10^{-33}$ and $10^{-32}~cm^2$. 
This paper should be considered to be a new idea which will need substantial new experimental input from all involved experiments.}


\maketitle
%
%________________________________________________________________

\section{Introduction} 

\begin{figure}[tbp]
\includegraphics[width=0.5\textwidth]{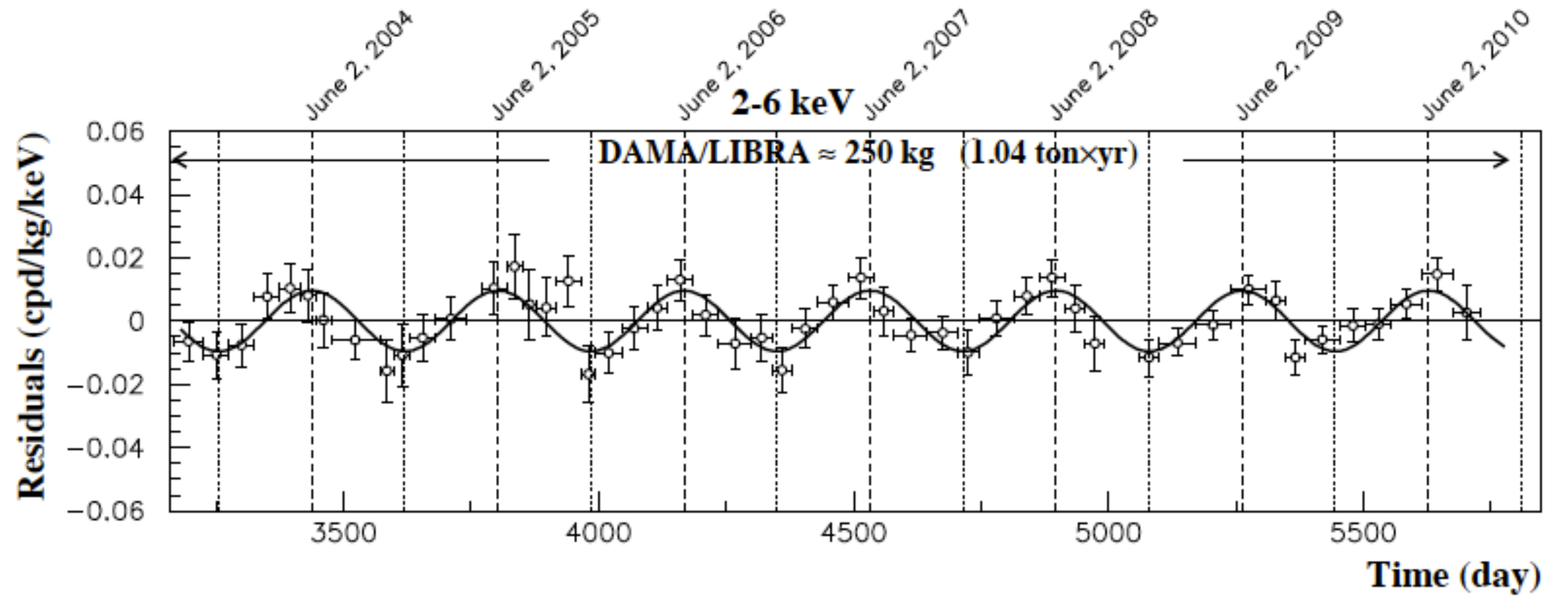}
\caption{DAMA data: residual rate of the single-hit scintillation events, measured by DAMA/LIBRA over 
six annual cycles in the (2-6) keV energy interval as a function of time [\cite{Bernabei_2013}]. The modulation amplitude is $\sim$0.01 counts/day/kg/keV.}
\label{fig:DAMA_results}
\end{figure}

The DAMA experiment observes a small, but clear, oscillatory signal. The signal is a peculiar modulation of the
counting rate of 2-6 keV single-hit events (i.e. those in which only one of the many detectors 
in the setup fires) satisfying all the many requirements of the Dark Matter annual modulation signature.  
The observed yearly modulation is in phase with the Earth's motion around the Sun. 
DAMA is the first experiment to observe a clear oscillatory signal [\cite{Bernabei_2002}] and [\cite{Bernabei_2013}], as shown on Fig.~\ref{fig:DAMA_results}.

\begin{figure}[tbp]
\includegraphics[width=0.4\textwidth]{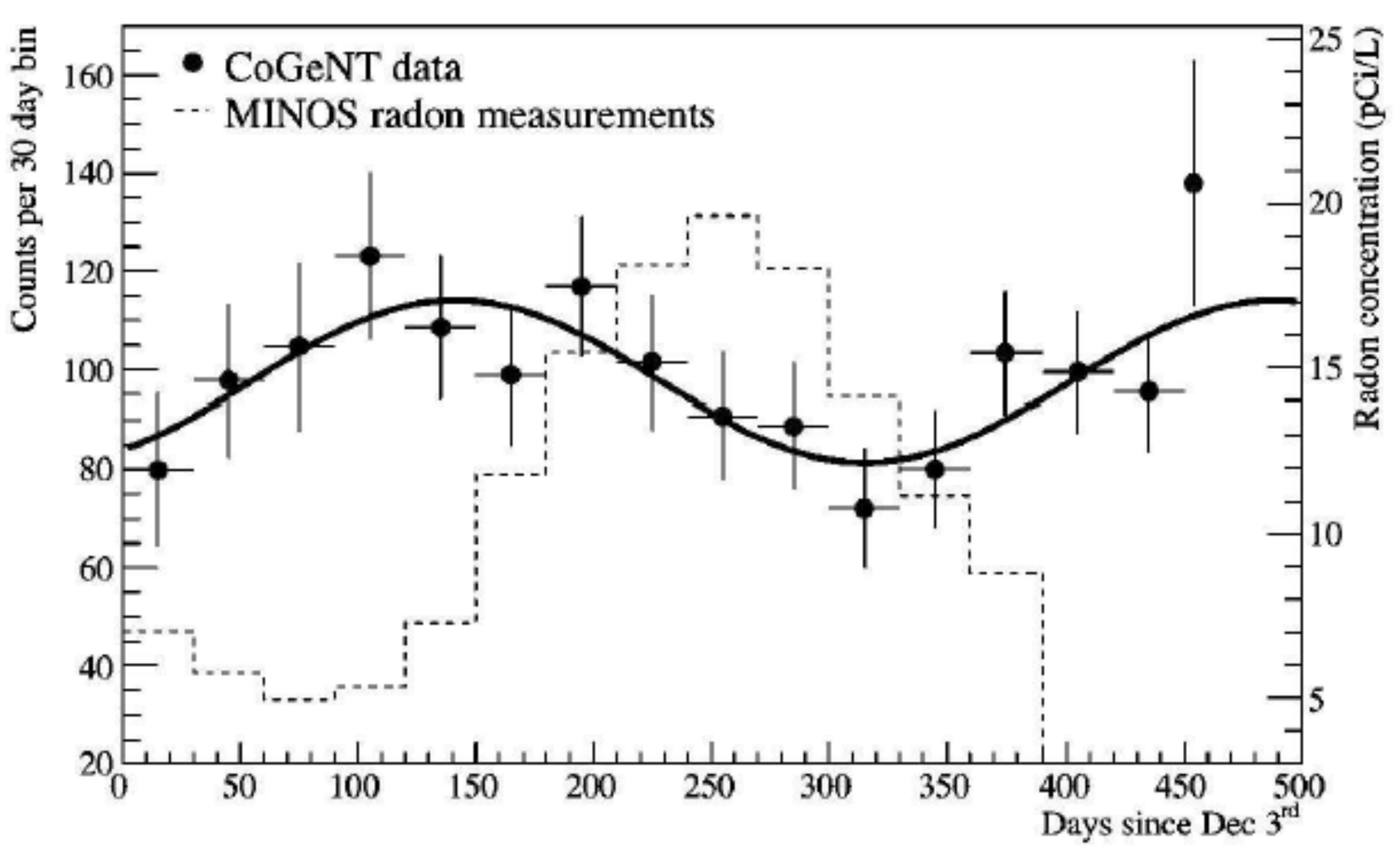} 
\caption{CoGeNT data: counts per 30-day interval from the 0.5-to-few keVee energy window [\cite{Aalseth_2013}]. 
The modulation amplitude is $\sim$0.3 counts/day/kg/keV.}
\label{fig:CoGeNT_results}
\end{figure}

The CoGeNT experiment (see Fig.~\ref{fig:CoGeNT_results}) also reports evidence of oscillation, and 
the CRESST-II and CDMS experiments also report hints of a signal. However, there is no hint of a signal from any LXe Dark Matter 
Search experiments, such as Xenon-10, Xenon-100 or LUX. 

\begin{figure}[tbp]
\includegraphics[width=0.45\textwidth]{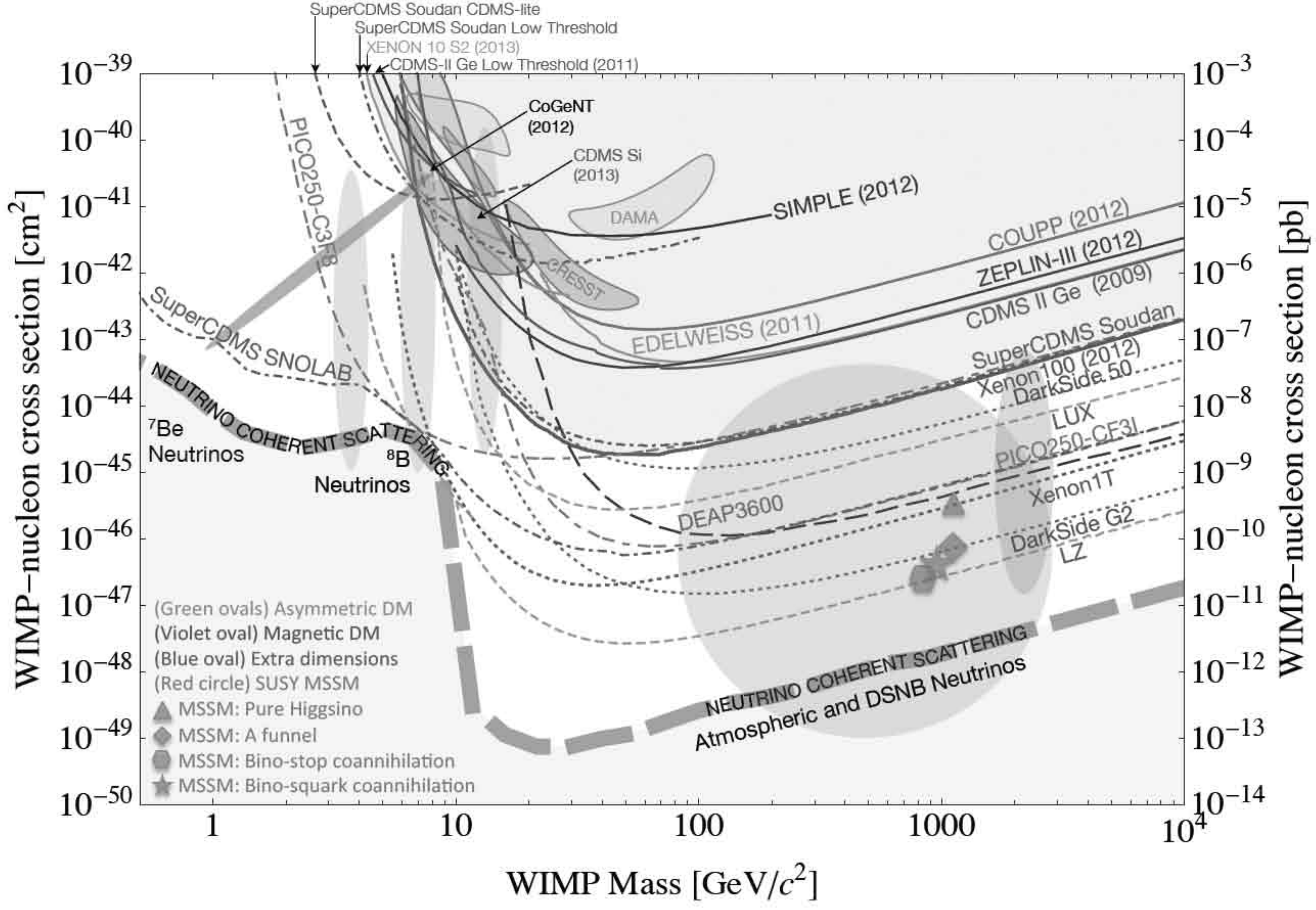}
\caption{The model-dependent status of cross section limits for the detection of the WIMP mass from various experiments (Snowmass, 2013). 
The graph also shows the low limit, below which experiments will have to deal with neutrino background. One can see clearly, that a 
WIMP mass below $\sim$5~GeV/$c^2$ would not be detected by most of the present experiments.}
\label{fig:WIMP_mass_status}
\end{figure}

Figure~\ref{fig:WIMP_mass_status}, which was shown at the 2013 Snowmass meeting, presents one of the possible model-dependent comparisons to the results of
dark matter experiments. The important point for this paper is that there is great opportunity to 
explore the cross section region of $10^{-30}-10^{-44}~cm^2$ if the mass is indeed low.

\begin{figure}[tbp]
\includegraphics[width=0.5\textwidth]{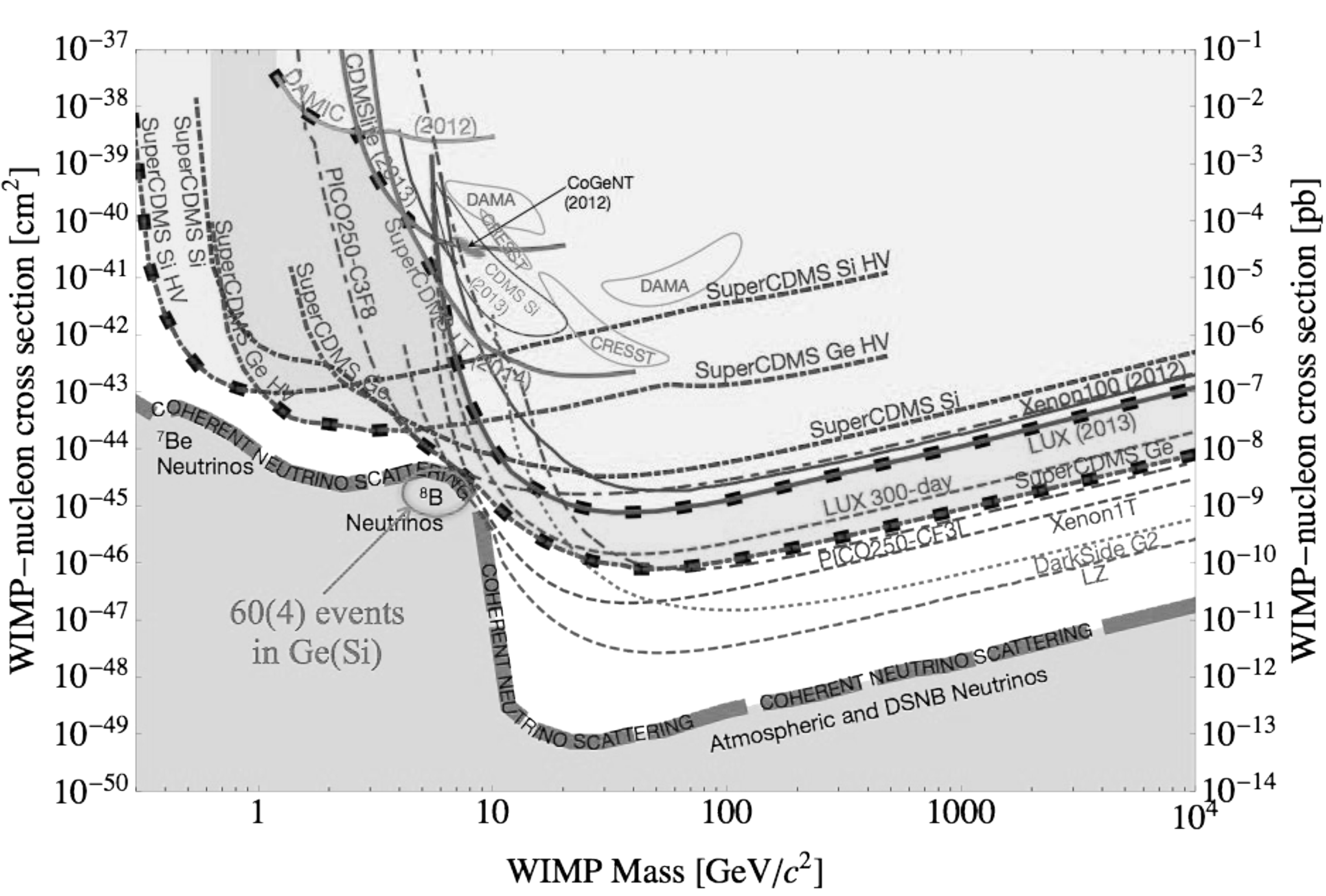}
\caption{As for Fig.~\ref{fig:WIMP_mass_status}, but the figure also shows the future reach of the CDMS experiment at low mass (presented 
by CDMS at Snowmass, 2013).$^1$}
\label{fig:CDMS_WIMP_mass_status}
\end{figure}

Figure~\ref{fig:CDMS_WIMP_mass_status} shows the CDMS future low mass limit reach\footnote[1]{B. Cabrera, 
private communication; figure presented at the Snowmass meeting, 2013 on behalf of the CDMS group.} for various new detector 
configurations and using new improved electronics. Clearly, CDMS aims to explore the low mass range, and so the idea presented
in this paper will be tested soon.

\section{A new idea to explain the DAMA data}

The author discusses a new idea, which explains the DAMA signal as being due to the scattering of a light Dark matter particle (1-to-few~GeV/$c^2$)
on a light nucleus; specifically on hydrogen present as a small OH or H$_{2}$O contamination in the NaI(Tl) crystals.

\begin{figure}[tbp]
\includegraphics[width=0.45\textwidth]{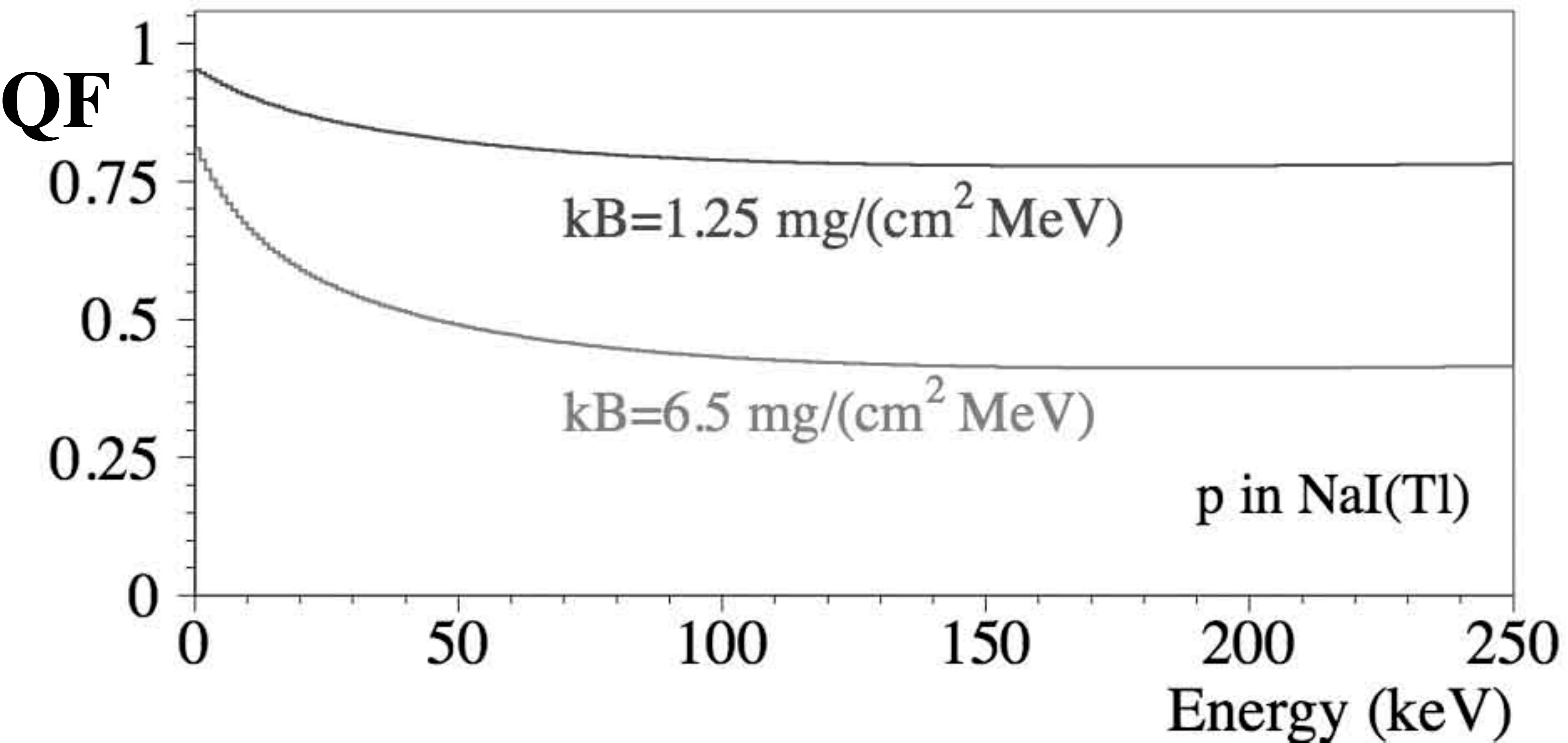}
\caption{The semi-empirical calculation$^5$ of the  quenching factor for proton recoils in the NaI(Tl) crystal, for two values of the Birks constant.
DAMA determined a low value of kB~=~1.25~mg/(cm$^2$~MeV) using alpha particles measured in their experiment; this in turn gives
a higher QF value for very small proton recoil energies.$^5$}
\label{fig:Quenching_factor}
\end{figure}

Table~\ref{table:1} shows the maximum calculated kinematical limit in terms of the nuclear recoil (keVnr) and the electron-equivalent 
recoil (keVee).\footnote[2]{It should be noted that the literature indicates quite a scatter in values of the Galactic escape velocity:
(a) 450 km/sec $<$ v$_{esc}$ $<$ 650 km/sec [\cite{Leonard_1990}], (b) 489 km/sec $<$ v$_{esc}$ $<$ 730 km/sec [\cite{Kochanek_1996}], and 
(c) 498 km/sec $<$ v$_{esc}$ $<$ 608 km/sec [\cite{Smith_2012}].} We point out that DAMA quotes the threshold in keVee units, so that 
one has to convert keVnr units to keVee units. One can easily calculate the keVnr-scale using two-body kinematics; to convert the keVnr-energy-scale
to the keVee-scale, one needs to know the quenching factor (QF) for a given nucleus, 
i.e., E$_{keVee}$ = QF~*~E$_{keVnr}$. The quenching factor depends on the recoil energy.
In case of NaI(Tl), for sodium recoil, I assume QF~$\sim$~0.25 with the understanding that the value could be lower.\footnote[3]{Some experiments
measured the quenching factor QF for sodium in NaI(Tl) close to 0.25-0.3, see for example [\cite{Gerbier_1999}] and [\cite{Tretyak_2014}]. 
However, according to private communication with J. Collar, some more recent measurements indicate values close to 0.1-0.15 for a recoil energy 
of $\sim$~10keVnr [\cite{Collar_2013}]; recentlt this has been measured independently at Duke University, with a similar result (soon to be published). 
However, as discussed in the paper, the QF factor depends on exact experimental conditions, i.e., it is not a universal constant [\cite{Tretyak_2014}].}
The proton quenching factor in NaI(Tl) has not been measured to our knowledge.\footnote[4]{Confirmed by private
communication with J. Collar, D. McKinsey and V.I. Tretyak.} Generally it is expected that the proton QF is considerably higher than for Na or I recoils, but 
not quite as high as 1.0. In the absence of any measurement, we follow the semi-empirical calculation of V. Tretyak [\cite{Tretyak_2014}], which 
is based on only one parameter, kB, the so-called Birks factor. The value of the kB factor depends [\cite{Tretyak_2014}] on many parameters, such 
as the detector material itself with all its impurities and dopants, the conditions of the experiment such as temperature, the 
signal integration time, etc. Ideally, the QF value should be
measured in the same experiment, and under exactly the same conditions as used for data taking. The kB factor could be determined by fitting 
experimental data obtained using one type of particle, for example $\alpha$ particles from an external source, and then using it to predict the
QF value for another type of particle, for example the proton, even in a different energy region. Figure~\ref{fig:Quenching_factor} shows
V. Tretyak's QF calculation\footnote[5]{Private communication with V.I. Tretyak. The same calculation was used in [\cite{Tretyak_2014}].} for the
proton QF for two values of the Birks constant, based on two extreme values obtained from different experiments; for kB constant at extreme values
between 1.25 and 6.5 mg/(cm$^2$~MeV), the proton QF is between 0.95 and 0.8 at $\sim$0 keV; for the DAMA case, where kB was determined to 
be 1.25~mg/(cm$^2$~MeV) using the $\alpha$ particle calibration, the proton QF is $\sim$~0.95 for very small proton recoil energies. 
I assume a value of QF~$\sim$~0.95 for the calculations of Table~\ref{table:1}. The conclusions would not change, however, if one 
would take QF $\sim$0.8 instead.

The DAMA experiment detection software threshold was 2~keVee during most of data taking [\cite{Bernabei_2008}], but it was recently 
lowered to 1~keVee [\cite{Bernabei_2012}] (while the hardware threshold was held at 1 photoelectron all along). 
Notice in Table~\ref{table:1} that the maximum electron-equivalent recoil energy for Na-nucleus is below
the new DAMA detection threshold unless the WIMP mass is above $\sim$4~GeV/$c^2$. On the other hand, it is possible to 
detect a larger WIMP mass than $\sim$1~GeV/$c^2$ for scattering off the H-nucleus even with the old DAMA threshold of 2~keVee, assuming that 
the proton QF is $\sim$0.95.

\begin{table}
\caption{Maximum calculated nuclear recoil energy E$_{keVnr}$ and maximum electron-equivalent
recoil energy E$_{keVee}$ as a function of WIMP mass, for two targets, hydrogen and sodium. It is 
assumed that the WIMP velocity relative to Earth is between 0 and the Galactic escape velocity 
of $\sim$650~km/sec, all modulated by the Earth's velocity of 230$\pm$30~km/sec in the Galactic coordinate system. 
In this table, for sodium recoil I used QF~=~0.25 [\cite{Gerbier_1999}], and for hydrogen QF~=~0.95.$^5$}

% title of Table
\label{table:1}                                                                                         % is used to refer this table in the text
\resizebox{9cm}{!} {
\begin{tabular}{c c c c}                                                                                % centered columns (4 columns)
\hline\hline                                                                                            % inserts double horizontal lines
WIMP mass [GeV/$c^2$] & Nucleus &  E$_{keVnr}$ [keV] & E$_{keVee}$ [keV] \\                             % table heading
\hline                                                                                                  % inserts single horizontal line
0.5 & H & 1.91 & 1.81 \\
1.0 & H & 4.30 & 4.08 \\
1.5 & H & 6.20 & 5.89 \\
2.0 & H & 7.65 & 7.26 \\
2.5 & H & 8.78 & 8.34 \\
3.0 & H & 9.68 & 9.20 \\
\hline
0.5 & Na & 0.19 & 0.05 \\
1.0 & Na & 0.73 & 0.18 \\
1.5 & Na & 1.57 & 0.39 \\
2.0 & Na & 2.67 & 0.67 \\
2.5 & Na & 4.00 & 0.60 \\
3.0 & Na & 5.53 & 1.38 \\
4.0 & Na & 9.07 & 2.26 \\
\hline\hline 

%inserts single line
\end{tabular}
}
\end{table}

\begin{figure}[tbp]
\includegraphics[width=0.5\textwidth]{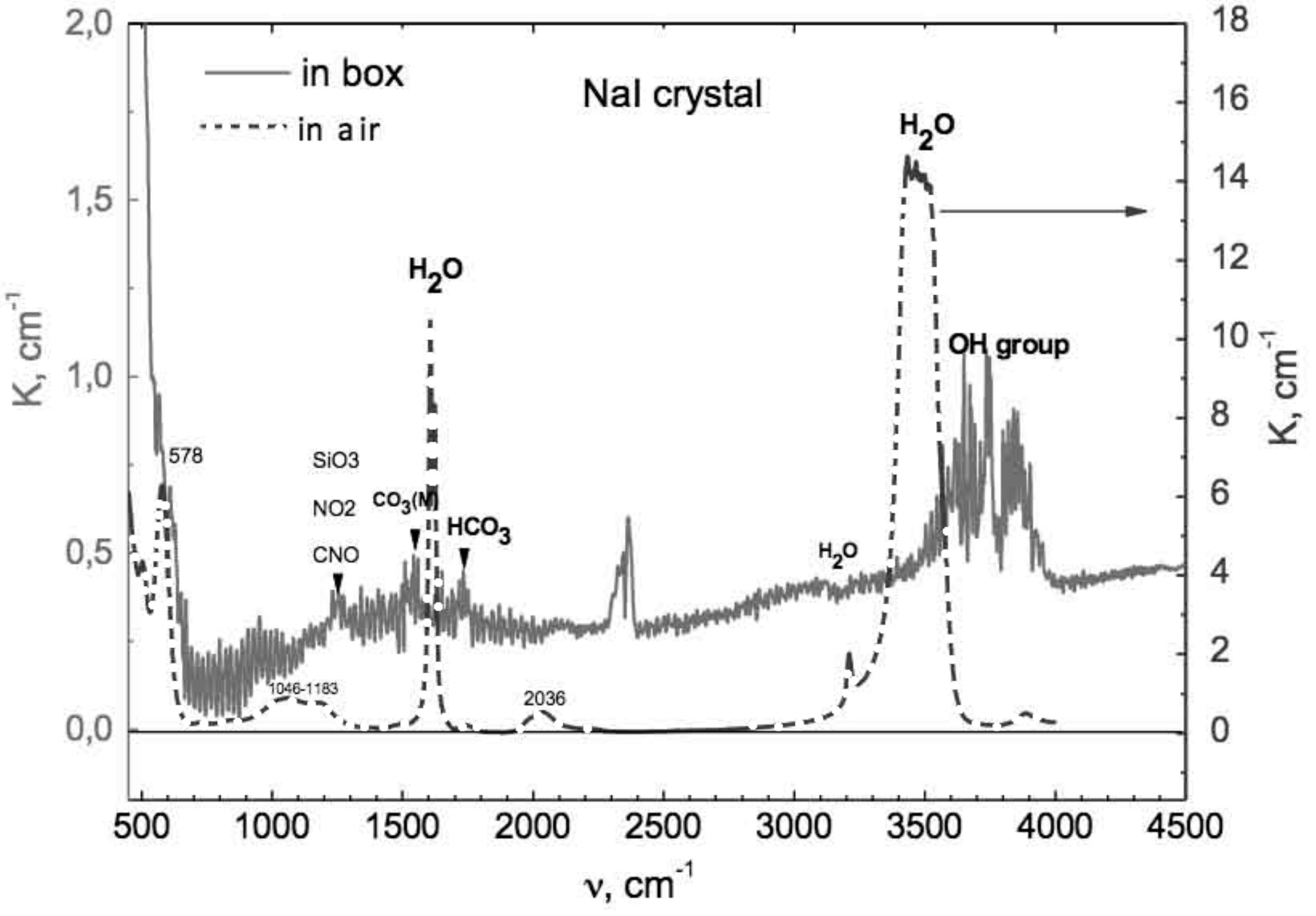}
\caption{FTIR spectrum of NaI(Tl) crystals showing the OH-contamination in a crystal located in a protective enclosure (solid curve) and the H$_{2}$O
contamination when exposed to air (dashed curve).$^6$}
\label{fig:OH_contamination}
\end{figure}

DAMA NaI(Tl) crystals are made of Ultra-Low-Background (ULB) materials to guarantee low radioactive background. While the DAMA experiment does 
quote very low radio-purity levels [\cite{Bernabei_2008}], it does not quote the H, OH and H$_{2}$O contamination at all.
Generally, the H-contamination is not quoted by crystal producers, since spectroscopic properties of NaI(Tl) crystals 
are not affected by it up to a level of $\sim$10~ppm. The OH-molecule represents a likely source of contamination of the primary NaI salt; its
removal requires prolonged pumping at elevated temperature [\cite{Kudin_2011}]. Figure~\ref{fig:OH_contamination}
shows typical OH-contamination levels in NaI(Tl) crystals\footnote[6]{N. Shiran, A. Gektin, Institute of Scintillating Materials,
Kharkov, Ukraine, private communication, 2014.}; the FTIR spectrum shows the OH-contamination for a crystal located in a protective 
enclosure (solid curve) just after its production, and the H$_{2}$O contamination when it was exposed to air (dashed curve). 
The exact level of the OH or  H$_{2}$O contamination in NaI(Tl) crystals is generally proprietary information, as told by Saint-Gobain Co.
However, we learned from the Hilger Co. that an OH-contamination at a level of 1-3~ppm is very likely. Generally, these companies keep the air 
humidity at less than $\sim$2$\%$ when machining a crystal and assembling an NaI(Tl) detector, although again this is proprietary 
knowledge. But, knowing how hard it is to keep humidity low in gases without appropriate filtration or without pumping while baking, it would seem to us that water 
contamination at a level of a few~ppm is easily possible. Once in the detector, each DAMA NaI(Tl) crystal is encapsulated in 
a copper housing and sealed for all its life. In addition, the detectors are continuously maintained in high purity nitrogen atmosphere. 
Therefore, if there is any H-contamination, it is probable that it would have occurred during previous manufacturing steps.  

While the possible presence of H/OH-contamination in the Ultra-Low-Background (ULB) NaI(Tl) DAMA detectors remains an open question, for 
our calculation we will assume, in an arbitrary way, that the H-contamination is at a level of 1 ppm. Once a better number is known, one
can easily scale our results. 

\section{Other Dark Matter search experiments}

If the proposed H-contamination mechanism works for the DAMA experiment, it must also work for CDMSlite [\cite{Agnese_2013}], CoGeNT [\cite{Aalseth_2013}], and
CRESST-II [\cite{Angloher_2011}]. The peak modulation amplitude for DAMA is $\sim$300~events/30 days, for CoGeNT it is $\sim$10~events/30 days; CDMSlite
measures $\sim$2.9 counts/keV/kg-day (no clear oscillation measured yet). 

Both the CDMSlite and CoGeNT experiments use ultra-pure 
Ge-crystals [\cite{Hansen_1982}]. Hydrogen is the only gas which has been successfully used for high-purity Ge-crystal growth. 
All commercial detector grade germanium is grown exclusively in hydrogen. Therefore, even the ultra-pure Ge has a 
H-contamination at a level of $\sim$2x$10^{15}$ atoms/cm$^{3}$, or $\sim$50~ppb [\cite{Hansen_1982}], and according to
P. Brink\footnote[7]{P.Brink, CDMS collaboration, private communication, 2014.} this is a reasonable estimate to assume.
However, we stress that no real measurement was made on the actual crystals from these two experiments, and so we encourage that this is done.
For our calculation we assume a level of $\sim$50 ppb.

The CRESST-II experiment uses CaWO$_{4}$ crystals of total weight $\sim$~10~kg. These have been prepared by the Czochralski growth 
method using the solid state reaction from CaCO$_{3}$ and WO$_{3}$ from Alfa-Aesar materials with purity levels of 99.999~$\%$ and 99.998~$\%$ 
respectively [\cite{Erb_2013}]. The paper does not mention the H-contamination level, but the purity level of the major ingredients used to form the crystal
is not very high, from our perspective. Reference [\cite{Li_2013}] describes another possible method to prepare CaWO$_{4}$ crystals, and 
they did measure a significant level of OH contamination using the FTIR method. However, we were not able to obtain information on the H-contamination 
for actual CRESST-II crystals, even after contacting members of that collaboration. Therefore, for our calculation we assume, in an 
arbitrary way, an H-contamination at a level of 100 ppb. Again, once a better value is known, our results can be easily scaled.

\section{Rate calculation}

A low mass WIMP of 1-to-few~GeV/$c^2$ represents a real experimental challenge at present for most experiments, as it requires a low mass
hydrogen target in order to be detectable. So far, existing experiments provide the hydrogen target only as a small "unwanted" contamination, which 
limits the measured rate. Alkali halide inorganic crystals, such as NaI(Tl) or CsI(Tl), may have a unique advantage over other 
methods in detecting a very-low-mass WIMP, if they have sufficiently high OH-contamination.

We calculate the rate as follows: 

\begin{equation}
Rate = \sigma_{WIMP} \cdot F_{WIMP} \cdot N_{target} \cdot V_{target} \cdot Time \cdot Det_{ef}
\end{equation}

where Det$_{ef}$ is the detection efficiency, $\sigma$$_{WIMP}$ is in [cm$^2$], F$_{WIMP}$ is flux in [cm$^{-2}$sec$^{-1}$],
N$_{target}$ is number of target particles in [cm$^{-3}$], V$_{target}$ is target volume in [cm$^{3}$] and Time is in [sec]. 
The WIMP-proton scattering cross section is not known at present. We will treat it as an independent variable. 
Table~\ref{table:2} lists the input numbers used in the calculation. 

\begin{figure}[tbp]
\includegraphics[width=0.5\textwidth]{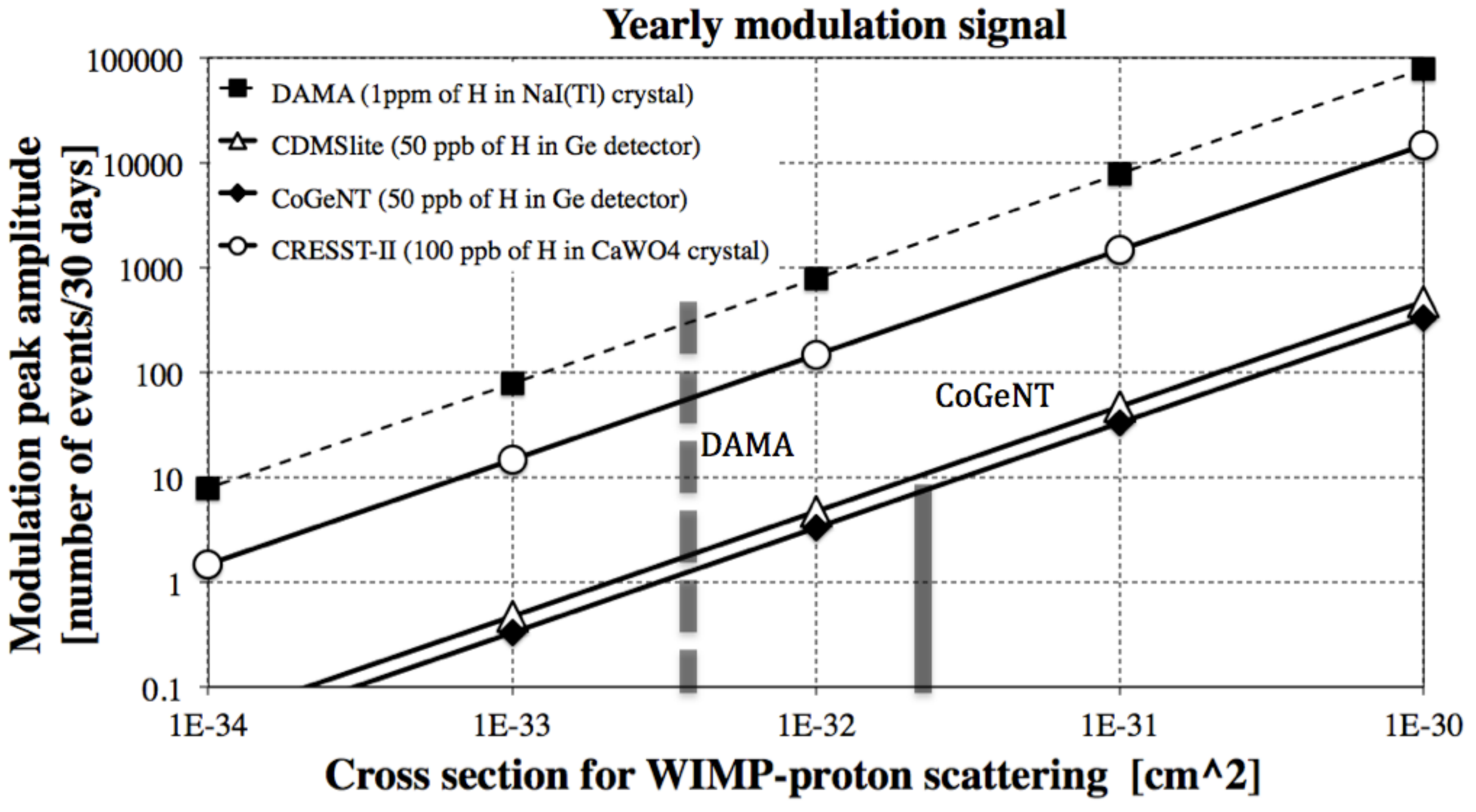}
\caption{This paper's prediction for the modulation peak rate in several experiments as a function of cross-section, assuming 
the H-contamination as indicated, WIMP-proton scattering hypothesis, Dark matter density $\sim$0.3~GeV/cm$^3$ in our nearby 
Universe [\cite{Catena_2011}]. The modulation amplitude peak was calculated from a differential rate based on the Earth's velocity
variation of $230\pm$30~km/sec in the Galactic coordinate system. The vertical lines indicate a cross-section corresponding to DAMA (dashed line)
and to CoGeNT (solid line) measured peak amplitude rates.}
\label{fig:DAMA_result_prediction}
\end{figure}

Figure~\ref{fig:DAMA_result_prediction} shows our prediction for the modulation peak rate in several experiments as a function of cross section, assuming
the H-contamination indicated in the figure, as a function of the WIMP-proton cross section.  
In order to obtain the DAMA modulation peak of $\sim$300 events/30 day period a cross section slightly lower than 
$\sim$10$^{-32}$~cm$^2$ is required. However, to explain the CoGeNT count rate, a slightly higher value of the cross section is required. It is 
therefore crucial that both experiments determine the H-contamination in their crystals.

\begin{table}
\caption{Input numbers used in the calculations for Fig.~\ref{fig:DAMA_result_prediction} and Table~\ref{table:3}.}
% title of Table
\label{table:2}                                                                                                 % is used to refer this table in the text
\resizebox{9cm}{!} {
\begin{tabular}{c c c c c}                                                                                      % centered columns (4 columns)
\hline\hline                                                                                                    % inserts double horizontal lines
Experiment & weight [kg] & H-contamination & Detection efficiency & Energy range [keVee] \\                     % table heading
\hline                                                                                                          % inserts single horizontal line
DAMA & 250 & 1 ppm & 0.6 & 1-6 \\
CoGeNT & 0.444 & 50 ppb & 1.0 & 0.5-3 \\
CDMSlite & 0.6 & 50 ppb & 1.0 & 0.5-3 \\
CRESST-II & 10 & 100 ppb & 1.0 & 0.5-3 \\
\hline\hline 

%inserts single line
\end{tabular}
}
\end{table}

Table~\ref{table:3} shows the number of events expected per 30 days in the modulation peak amplitude for four experiments.

\begin{table}
\caption{Calculated number of events expected over 30 days in the peak of the modulation, for DAMA, CoGeNT, 
CDMSlite and CRESST-II experiments using the H-contamination model.}
% title of Table
\label{table:3}                                                                                                 % is used to refer this table in the text
\resizebox{9cm}{!} {
\begin{tabular}{c c c c c}                                                                                        % centered columns (4 columns)
\hline\hline                                                                                                    % inserts double horizontal lines
Cross section [cm$^2$] & DAMA & CoGeNT & CDMSlite & CRESST-II \\                                                            % table heading
\hline                                                                                                          % inserts single horizontal line
10$^{-32}$ & 794 & 3.25 & 4.69 & 146.6 \\
10$^{-33}$ & 79.4 & 0.33 & 0.47 & 14.7 \\
10$^{-34}$ & 7.9 & 0.03 & 0.047 & 1.47 \\
\hline\hline 

%inserts single line
\end{tabular}
}
\end{table}

In order to make a quantitative comparison to the proposed model, each of the four experiments should
perform an H-contamination analysis of the crystals used in that experiment; DAMA should also find out if there is a correlation 
between rate/crystal and the H-contamination in each crystal (if indeed there is a variation in rate/crystal).

The calculation in Figure~\ref{fig:DAMA_result_prediction} is made simply to get a feel for the problem, and to provide motivation to measure 
the hydrogen contamination in the existing crystals. The exact tuning of the dark matter particle mass 
between 1 and a few~GeV/$c^2$ should be done with a proper fit, which includes a realistic simulation of the dark matter velocity
distribution in the Galaxy, the cross section, and the H-contamination. An example of such a calculation is found in Ref. [\cite{Profumo_2013}], and
this disagreed with our hypothesis of hydrogen contamination, as presented in our original paper [\cite{Vavra_Jan_2014}]. 
However, these authors assumed a Maxwellian distribution of dark matter velocities in our Galaxy. 
We would argue that the dark matter velocity distribution may have two components in reality, one component 
representing the Maxwellian distribution, and the second component representing a direct external non-Maxwellian flow distribution, as 
all galaxies may be interconnected by streams of dark matter with velocities above the Galaxy escape velocity. The direct flow component may 
enhance the reach to a lower WIMP mass range than that which is shown in Table~\ref{table:1}. 
This problem should clearly be studied in more detail.

Reference [\cite{Profumo_2013}] also pointed out a possible conflict with a model of overheating of the Earth's atmosphere if low 
mass dark matter particles have too large a scattering cross section [\cite{Mack_2007}]. However, from Ref. [\cite{Mack_2007}] we 
conclude that we are still below their quoted overheating limits, for the proposed dark matter masses near $\sim$1~GeV/$c^2$ and 
a cross section below $\sim$10$^{-32}$~cm$^2$. However, if the local dark matter density increased by a factor of $\sim$~10, or if 
the WIMP-proton scattering cross section is larger, the overheating described in Ref. [\cite{Mack_2007}], might occur.

The LXe and LAr experiments do not have hydrogen contamination, and therefore any DM signal detected by these methods would directly contradict the proposed model.

\section{Comparison with LHC experiments}

\begin{figure}[tbp]
\includegraphics[width=0.5\textwidth]{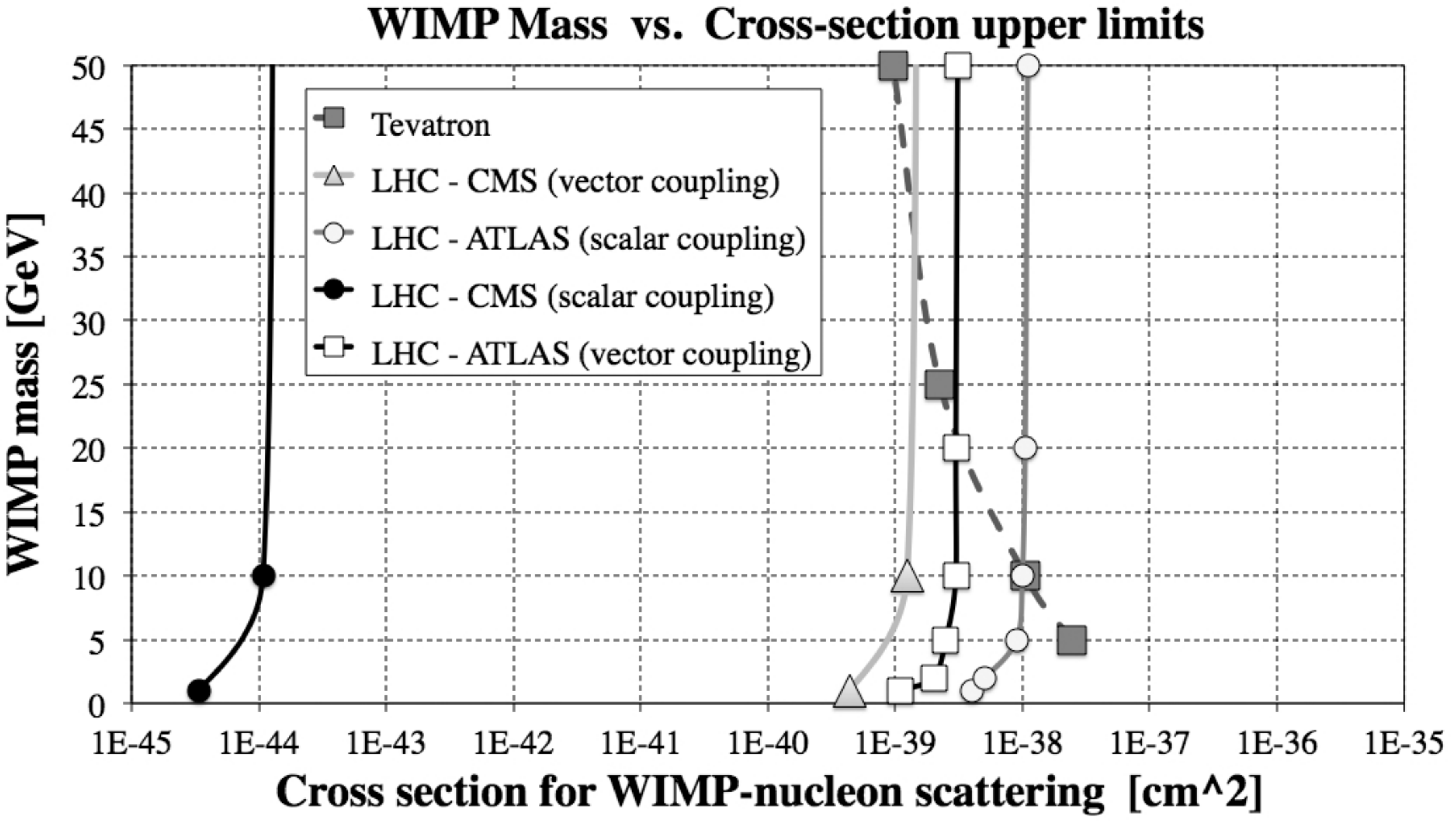}
\caption{Upper limits of WIMP-nucleon cross-sections as a function of the WIMP mass, plotted for Tevatron [\cite{Beltran_2010}] and 
LHC [\cite{ATLAS_2013}],CMS [\cite{CMS_2013}].
There is a large range of possible values of cross-section limits for $\sim$1~GeV/$c^2$ WIMP.}
\label{fig:Tevatron_and_LHC_results}
\end{figure}

It is interesting to compare our prediction for the DAMA WIMP-nucleon cross section range with upper limits obtained by Tevatron and LHC experiments, where 
the WIMP is measured in "$\overline{WIMP}$~+~$WIMP$~+~jets" events [\cite{Beltran_2010}]. This is shown in Figure~\ref{fig:Tevatron_and_LHC_results}.
It is clear that the accelerator-based cross section upper limits are lower than values obtained from the H-contamination model.
However the limits of collider experiments are derived with an effective operator approach, and direct comparison with our model 
has some unknown model dependency.

\section{A new proposal for future experiments}

One could attempt to detect diatomic molecular vibration, excited by gentle WIMP-proton scattering, and hence further reduce the threshold 
on the WIMP mass [\cite{Vavra_Feb_2014}]. To excite such vibrations, a very small energy deposit at a level of 1.8-4.3~eV is needed. 
Some fraction of de-excitations goes in the visible range, so it should be possible to detect it with the Bialkali photocathode; an IR-responsive photocathode 
would provide an even-more-efficient method. To do this, one could use, for example, so called "wet" fused silica, which 
has $\sim$1000~ppm of OH-contamination by design. Or, one could use simply pure water, and "tune" for the OH-absorption lines. 
This avenue is presently not pursued in underground experiments 
because of high single photo-electron noise. It could be considered in accelerator-triggered beam dump experiments searching 
for dark matter production. Pilot studies of this idea are in progress.

\section{Conclusion}

This paper suggests that the measured oscillation in the DAMA experiment may be caused by scattering of a light mass WIMP (mass between 
$\sim$1 and a few~GeV/$c^2$) on a hydrogen nucleus, present as a consequence of H-contamination in the NaI(Tl) crystals.

For our choice of parameters for the H-contamination model, and taking into account the DAMA modulation peak rate of $\sim$0.01 cpd/kg/keV, the 
observed behavior can be explained with a WIMP-proton scattering cross section between 10$^{-33}$~cm$^2$ and 10$^{-32}$~cm$^2$ 
(see Fig.~\ref{fig:DAMA_result_prediction}).

As the H-contamination may vary from crystal-to-crystal in the DAMA experiment. Crystal-dependent rates may be observed. This should be investigated.  

We suggest that the DAMA, CDMS, CoGeNT, and CRESST-II collaborations perform an FTIR analysis on their crystals in order to measure the 
level of H-contamination. This will prove or disprove our claim that a single WIMP-proton scattering cross section can explain all of the data. 

LXe or LAr experiments do not have hydrogen contamination, and therefore any DM signal detected by these methods would directly contradict 
the model proposed in this paper.

While it is somewhat puzzling that the idea presented in this paper requires a very small dark matter mass with a cross section many orders of 
magnitude larger than that for a particle on the mass scale of the W and Z bosons, there is nothing \textit{a priori} which, to our knowledge, prevents 
this from occurring, and so we suggest that the idea be tested very carefully.

\section{Acknowledgements}

 I also would like to thank CDMS people for discussing issues about the dark matter detection, especially Prof. B. Cabrera, S. Yellin, 
 P. Brink, R. Partridge, G. Godfrey, and M. Kelsey. I also thank Farinaldo Da Silva Queiroz and Chris Kelso for discussions regarding 
 reconciliation of ideas of this paper and their publication [\cite{Profumo_2013}]. Comments of Prof. C. Galbiati encouraging to 
 investigate this problem carefully are also appreciated. I would like thank Prof. E. Nappi for showing interest in the idea, sharing 
 it with Prof. C. Broggini, and asking very good questions about early version of the paper. I would like to thanks 
 Prof. E.V. Kolb for providing several references, which allowed me to enter the up-to-date status of the collider WIMP searches.

\end{document}